# THE VALUE OF USING BIG DATA TECHNOLOGIES IN COMPUTATIONAL SOCIAL SCIENCE

Eugene Ch'ng
School of Computer Science, University of Nottingham Ningbo China
199 Taikang East Road, Zhejiang Ningbo 315100, China
00 86 (0) 574-88183049
eugene.chng@nottingham.edu.cn

## ABSTRACT
The discovery of phenomena in social networks has prompted renewed interests in the field. Data in social networks however can be massive, requiring scalable Big Data architecture. Conversely, research in Big Data needs the volume and velocity of social media data for testing its scalability. Not only so, appropriate data processing and mining of acquired datasets involve complex issues in the variety, veracity, and variability of the data, after which visualisation must occur before we can see fruition in our efforts. This article presents topical, multimodal, and longitudinal social media datasets from the integration of various scalable open source technologies. The article details the process that led to the discovery of social information landscapes within the Twitter social network, highlighting the experience of dealing with social media datasets, using a funneling approach so that data becomes manageable. The article demonstrated the feasibility and value of using scalable open source technologies for acquiring massive, connected datasets for research in the social sciences.

## 1. INTRODUCTION
Social media and the underlying networks are increasingly important in the academia, particularly in the social sciences. Manual approaches that predate automated data collection and analysis however, are still being used for studying these new forms of societal expressions. As manual approaches can capture only a tiny fraction of the data, they may no longer be suitable as the context of a social network spans broad spatial-temporal landscapes. The larger context of societal networks require automated methods for acquiring and processing unstructured data that are longitudinal, relational and multi-modal. The study of social media data therefore remains a challenge, as the quantity, magnitude, and complexity associated with the volume, velocity and variety of data can be difficult to manage. Traditional approaches could become extremely tedious when data is big. 21st century social science data and the underlying networks is therefore a Big Data problem. Massive quantities of information generated by people are being tapped by diverse groups with the hope that it will answer questions in their disciplines. Regardless of which disciplines, the fact is that 'these massive amount of information can be tracked and measured with unprecedented fidelity' [1] in the Big Data context. A straightforward search in Google Trends on 'Big Data' as compared to other trending keywords shows unprecedented interest as compared to other trending keywords. This may indicate that the need for Big Data pervades disciplines that make use of data. Data can be 'Big' in different ways, and is observed by Manovich [2] as datasets that are sufficiently large to require supercomputers. However, Big Data is not only characterised by its size, but by its relationality with other data [3] – "Big Data is fundamentally networked". Social media data if captured from various social media sites, potentially faces challenging issues related to all the 7 Vs (see Figure 1) due to unpredictability.

This article focuses on a data funneling process that makes large collections of topical, multimodal, and longitudinal Twitter data manageable and meaningful to the social sciences using scalable open source technology. The approach led to the discovery of interactions and communities within Twitter, which may lead to knowledge in large-scale online community-level activities, crowd-sourced sentiments and predictions.

## 2. BACKGROUND
80% of the world's unstructured data is untapped by the academia. These data are important to the 21st century quest for knowledge and can be acquired via state-of-the-art open source asynchronous Web technologies, stored and processed using distributed databases, mined via data mining and machine learning techniques, and rendered and visualised in multicore computer graphics workstations. The volume, velocity and variety users data can be posited within the Big Data domain and therefore, technological infrastructure is needed for presenting the data in a meaningful way. Data however, must be connected and structured in order for patterns to emerge. The value of Big Data "comes from patterns that can be derived by making connections between pieces of data, about an individual, about individuals in relation to others, about groups of people, or simply about the structure of information itself" [3]. When data becomes big and highly relational, it could potentially transform grounded theory, or possibly the understanding of it [1].

A very important property within social media is the connectedness of entities. There are hidden structures as a result of purposes behind actors, individual psychological states, their comments, interactions such as conversations, and shared semantics. Connecting these entities is a crucial step towards revealing the larger context of a person or group. These data can potentially uncover activities, communities and how information flows. These are of value to both the academia and the industry. Twitter, one of the largest social media is one such service that has become of high interest to researchers. Who would have thought that a simple service with 140-character message limit could be used for research in diverse fields? A handful of messages on the computer screens are manageable to a researcher.

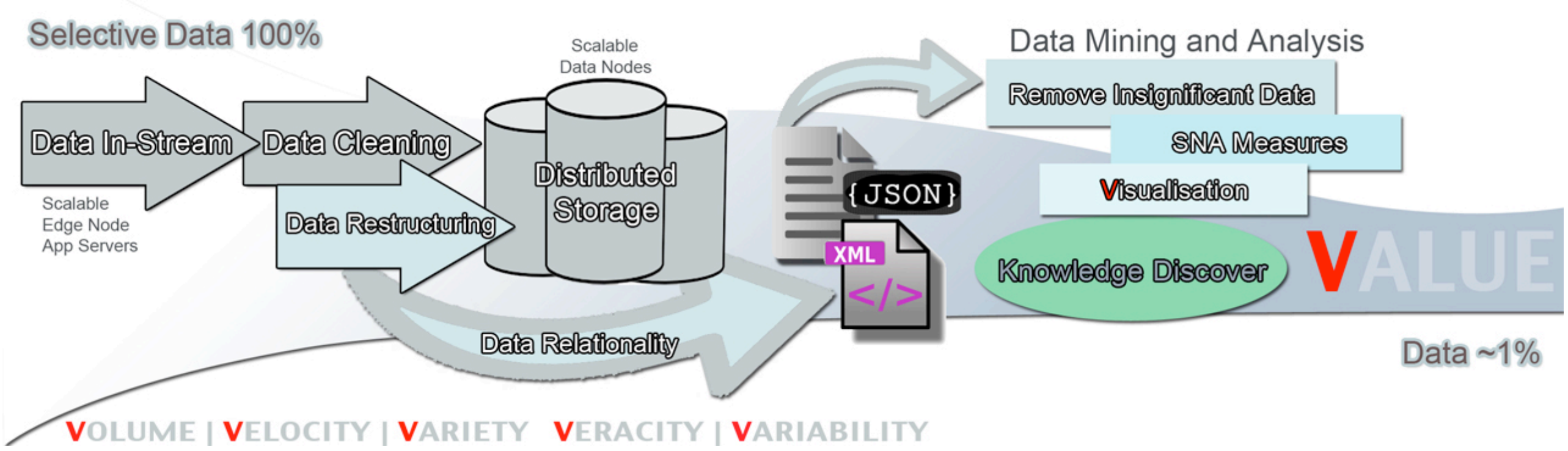

**Figure 1. A pipeline of process for making data manageable, relational and therefore valuable.**

However, trending and longitudinal tweets fall in the domain of Big Data research (see Figure 2 for peaks and valleys in data volume and velocity). The multimodality of social network data is also a growing issue. Whilst unimodal (relationships amongs friends, etc) research in Twitter networks exist, it has become apparent that multimodal networks between people, objects, tasks, etc are much more useful and of a broader scope [6].

Big Data offers hope in computational social science. The capacity to automate the collection and processing of social data is something that Big Data research can offer. However, dealing with Big Data is a major issue. Data may be big, but what makes Big Data valuable is the relationality that makes the discovery of patterns and hidden structures possible. This allows data consumers to discover knowledge and make informed decisions. Here, the funneling process that makes data manageable plays an important role.

## 3. METHODS

This section describes the funneling approach that this article presents for making multimodal, longitudinal social media data meaningful.

### 3.1 Data Manageability and Value

Data manageability plays an important role in making Big Data useful. How much of data can become useful depends on how well we can mine data for patterns. A collection of data may be in the tera or petabytes, within which perhaps only 1% (for example) may be useful. But without the 99%, that 1% may never be found. It is therefore important to collect and keep all potentially useful data, and, from those collections, conduct data mining in order to find the 1%. Such a concept involves a funneling process (Figure 1 is self-explanatory).

### 3.2 A Scalable Open Source Architecture

Open source libraries and Web technologies have been well tested and have such high efficiency that large corporations adopt them (e.g., PayPal, Yahoo and LinkedIn's use of NodeJS, FourSquare's use of MongoDB). Developing within these environments allows highly scalable and efficient applications.

There are two initial issues in Big Data social media research – velocity and the volume of data. In dealing with velocity, it is important to anticipate that longitudinal datasets captured in social media will be erratic and unpredictable in terms of volume and velocity (see Figure 2). A non-blocking I/O model that is data-intensive with the ability to push and pull data from multiple data sources in real-time is needed. As for the volume of data, the big truth about Big Data is that "it's easier to get the data in than out…" [6] from relational database management systems. The storage of Big Data in the tera- to petabytes require a format that can be stored, but accessed quickly and processed on the fly in real-time. As raw data from social media will be inconsistent, unstructured and variable, indexing may be difficult. As noted by Jacobs, "To achieve acceptable performance for highly order-dependent queries on truly large data, one must be willing to consider abandoning the purely relational database model". NoSQL (key-value pairs) databases that scales massively is necessary. One of the most important aspects of statistical correlation is the storage of as much data as a project can possible obtain. Storing all data for filtering later is a better approach than not having the full amount of data, for data lost means opportunity lost forever. This is important as social media services conduct routine archiving and removal of old data.

### 3.3 Architecture and Distributed Data

The scalable software architecture developed here are installed on a total of 16 Linux 64-bit Ubuntu (12.04) Virtual Machines (VM) within four (4x) Dell PowerEdge C6100, each with 2x Intel Xeon X5660 Processor (2.80GHz, 12M Cache, 6.40 GT/s QPI, Turbo, HT), 1333MHz Max Memory, 48GB Memory for 2CPU (6x8GB Dual Rank RDIMMs) 1333MHz, 250GB SATA 7.2k 3.5" HDD. This was found to be sufficient for the data captured, stored in the distributed database and processed in real-time, as data with irregular volume and velocity streams in. The setup below is sufficient for all the data gathered here. Replicating the scalable architecture on more VMs will be straightforward:

1. Each server runs a Node.js "edge node" application within one VM (1x core, 8GB RAM and 60GB HDD) that streams data. Node.JS works in parallel with other Node.JS instances for data streaming scalability.
2. The MongoDB server runs on VM (Quad-cores, 8GB RAM and 60GB HDD).
3. 3x MongoDB config servers running Ubuntu 13.04 "Saucy Salamander", a lightweight variant in the VM (Dual-core, 4GB RAM and 20GB HDD).
4. 3x Shards running on Ubuntu 13.04 "Saucy Salamander" in VMs (Dual-core, 4GB RAM and 20GB HDD). This is scalable to more shards as data size increases.

Data is collected using the Twitter API with keywords as topical filters. This is the initial data funneling process, separating noise from useful data. Data is distributed over three MongoDB shards with JSON key-pair values. The stored data is cleaned and includes socially relevant information such as userid, tweet, description, followers count, friend count, favourites count, time-zones, statuses count, geolocation, place, country, etc. Due to the uncertainty of tweets from very different continents, a cleaning function using regular expressions removes unwanted characters that can interfere with the GEXF XML structure (e.g., “\r\n\t:;,&'\<>.”) and replaces it with character entities (e.g., “& ' " < >”).

## 3.4 Relationality of Data Entities

Mapping multimodal activities within Twitter is much more valuable than the common follower-followee network, which has low activities as most users are inactive. The datasets here are actual interaction networks within Twitter.

An algorithm in the application parses each tweet and stores the actor “@UserA” and mentions of other actors, e.g., TweetA=“@UserB @UserC” into a Node array. Each tweet is stored as a TweetNode. Any mentions become a connection with @UserA→{TweetA}, @UserA→{@UserB, @UserC}, @TweetA→{@UserB, @UserC}, including (@UserA node connecting 3 nodes, 2 of which are users). Such a simple yet effective multimodal mapping technique has not been attempted before but could reveal various community related activities and information flow. Whilst data can be in any of the networks-based format (GraphML, XGMML, etc), GEXF is used for storing the network structure as it preceded other formats in terms of robustness and flexibility. GEXF provides a way to visualise network dynamics and evolution.

## 3.5 Visualising Processed Data

The GEXF datasets were ported into Gephi, a graph visualisation package for processing. Network topology filters were applied in order to reduce the number of nodes and edges, removing entities (retweets, etc.) without affecting important network hubs (>n in/out-degree) in the datasets (Figure 2 N/E labels with filter indicators ‘<#’). At this stage, the data is structured, and of a significantly smaller size due to the funneling process. An Intel i7-990x 6-core 12 threads workstation with 24GB RAM and Zotac GeForce GTX 560Ti CUDA graphics card were used for processing and visualisation. Centrality measures such as Degree, Betweenness, Closeness, Eigenvector were applied to the network in order to discover important egos before clustering algorithms (Force Atlas 1 and 2) were applied so that nodes with a higher number of interactions are closer to each other.

# 4. RESULTS

This section demonstrates some of the discoveries using the approach covered in the methodology.

## 4.1 Activity Signatures and Ego Centralities

The nodes and edges shown at the middle and bottom row of Figure 2 are processed using the funneling approach. Larger nodes have high degree centralities, the colour intensity of the node indicates strong Betweenness centrality, the label colour from black (low) to green/purple (high) has higher Closeness centrality. Each graph in the figure is a 5-hour dataset from a continuum of longitudinal datasets (a single point in the graph, top row).

The different signatures of the filtered graphs (bottom row) are due to the nature of the activities, the background of the topic and the actor intentions. Actors who interacted more are closest. They formed natural clusters. The #FreeJahar dataset has heightened activities (shades of purple labels) on the top part of the landscape. In the #PRU graph, two large political parties contended during the Malaysian general election bridged by activists, the opposition leader @AnwarIbrahim is flanked by media channels and have large Degree and Betweenness centralities, an indication of intense activities. In the controversial #NSA news, within the graph, a Guardian correspondent is highly connected, with clusters of activities below, note that #KatyPerry is above in a separate but linked community. The glamorous birth of the #RoyalBaby was mentioned by celebrities (e.g., @selenagomez, etc) with pockets of discussions throughout. The reawakening of the Madeleine #McCann kidnapping case due to new evidence led to separate clusters of varying degree of intense conversations. The MH370 dataset contains two separate clusters

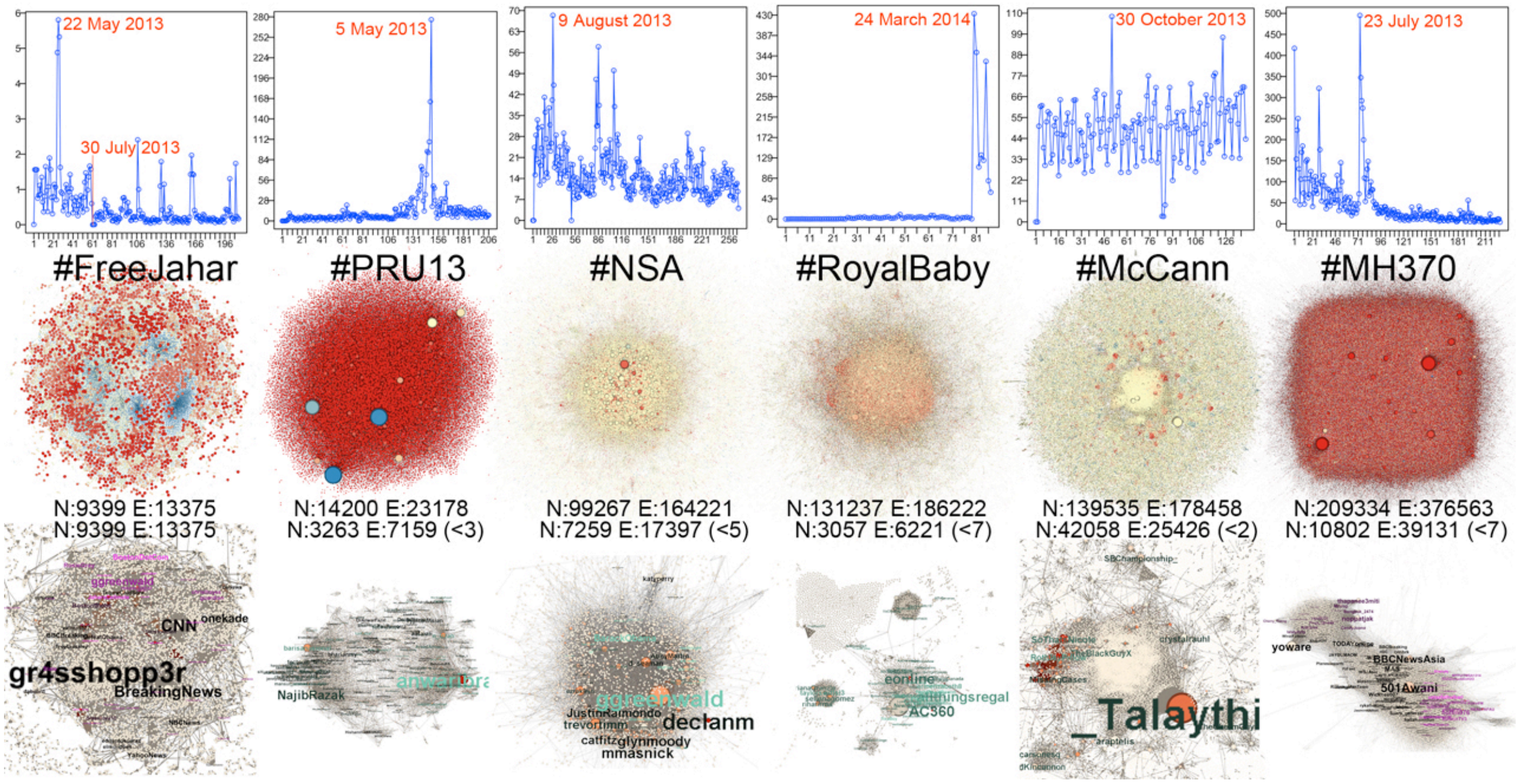

**Figure 2. Trends and viral outbursts of topical Twitter data (top graph). Each point is a 5 hour recording of Twitter activities. The #MH370 datasets have over 200 points for example. Each interaction network here (mid and bottom) is picked from a single**

of important activists. The brief overview of the datasets raises some interesting questions. For example, why are there many more distinct clusters of activities within the #McCann graph as compared to the others? Is this an indication that there are many kidnapping related experience in the clusters? Whilst the answers to these questions are of high academic interests, they are beyond the scope of this paper. What is more important is the byproducts of the Big Data software architecture, demonstrating how managing data through the funneling approach leads to the discovery of community activities in Twitter.

## 4.2 Discovering Communities

We now focus our attention on a single dataset (#FreeJahar in Figure 3). The majority of tweets outside of the cluster in the figure are retweets (RT). This gives rise to another phase in the funneling approach that segregates tweets with the keyword "RT" from the other tweets by visualising RT in black. The #FreeJahar dataset is associated with a large number of teenage girls who called for the freedom of the younger Boston bombing suspect (nicknamed Jahar) only because they believed he is "too beautiful to be a terrorist" [7]. The younger Boston suspect has since become a teen heartthrob as thousands of girls express their love for the bomber in online forums.

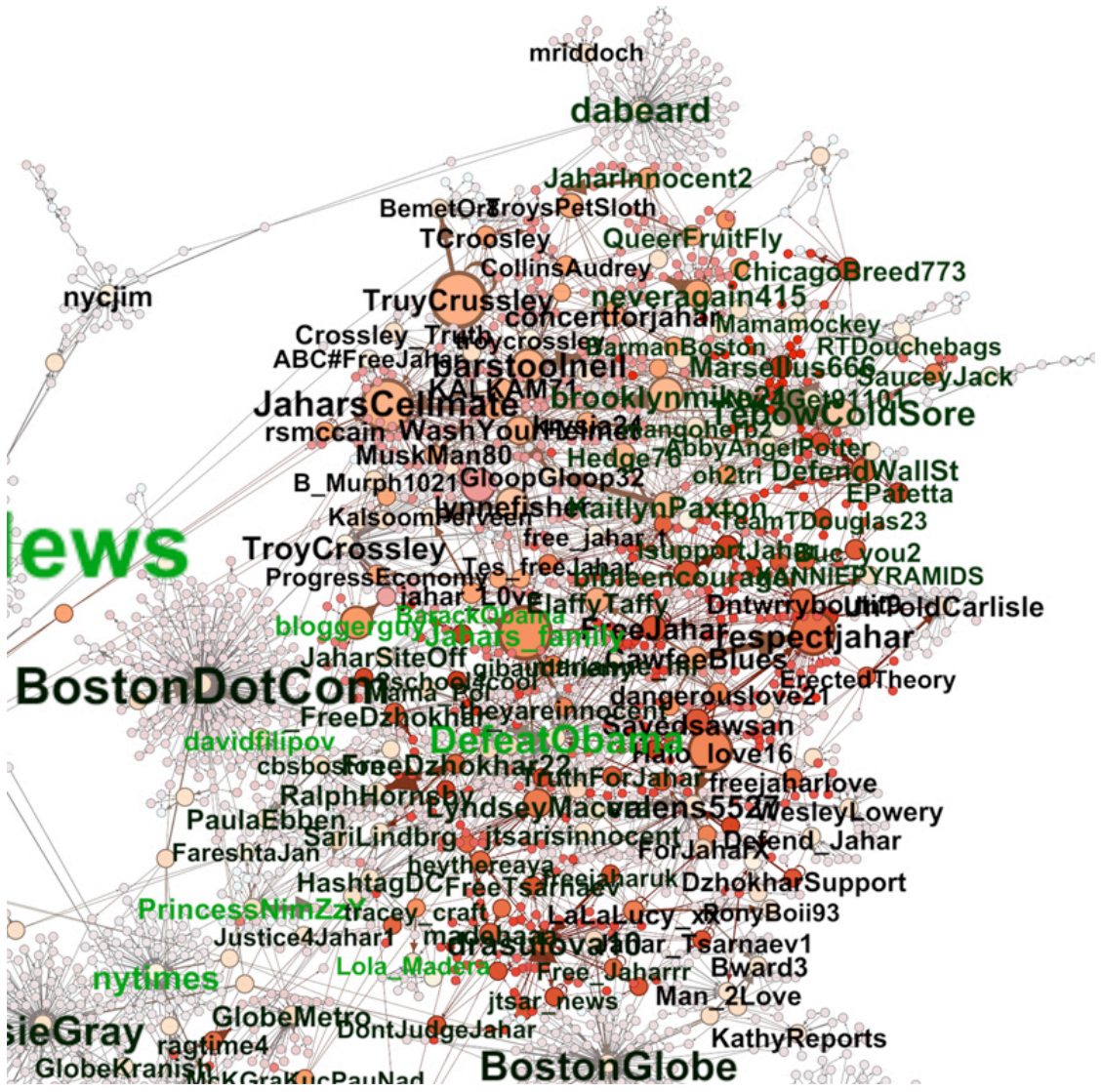

**Figure 3. A dataset from the #FreeJahar hashtag focusing on the heightened activities showing persistence of community activities, interaction boundaries and membership symbols.**

This dataset present a basis for studying the formation and decline of a community. Twitter is unlike other services such as Facebook, etc., that allows a formal formation of online communities as it was originally created as a messaging service. How was it possible that communities formed within a social media service that does not support formal group formations? If a cluster does form, can it be classified as a community at all? Traditionally, the concept of a community involves territorial boundaries, but the modern notion of community is better defined by the nature of relationships rather than on geographical proximity [8]. The #FreeJahar dataset conformed to a "sense of community" as defined by McMillan and Chavis [9]. Data analysis shows that community boundary exists, which segregates the in-groups from the out-groups. The community also adopts some types of symbols, and members exert influence in the group. Furthermore, community persists across the data continuum, with a bottom-up organisation of the community. In social science research, formulating hypotheses and conducting interpretations are definitely necessary for revealing the phenomena in Figure 3 in more detail. However, since the scope and aims of this paper is to demonstrate the feasibility of using open source Big Data technology, a description of the findings here will suffice.

# 5. CONCLUSION

21st century social science data is a Big Data issue and presents great technological challenges. Unlike machines and sensor devices which have very predictable data, user generated contents are unpredictable, culture-influenced, event-driven and topic-based. The content of social media could be from very different psychological states and social context. The Big Data architecture and the funneling method that deals with the final manageability of data are important here. The experience of dealing with social media data is a good one as it challenges traditional methodology, and possibly theoretical foundations in the social sciences. It would be interesting to see where social science theories would stand if and when data becomes Big and all-encompassing. Work is currently underway to analyse the datasets associated with the six hashtags by connecting theories with real-world data.

# 6. ACKNOWLEDGMENTS

The author acknowledges the financial support from the International Doctoral Innovation Centre, Ningbo Education Bureau, Ningbo Science and Technology Bureau, China's MoST and The University of Nottingham